# ORCHESTRATION PARADOXES IN NATIONAL QUANTUM COMPUTING INNOVATION ECOSYSTEMS

Jori Taipale, Olli Tyrväinen, Tuure Tuunanen, Joel Mero, Teiko Heinosaari

Abstract

*Effective orchestration is a critical driver of success in quantum computing innovation (QCI) ecosystems. Heterogeneous actor goals, roles, and power relations, however, produce tensions that confront orchestrators with paradoxical situations in which they must navigate trade-offs between competing demands. To orchestrate an ecosystem effectively, these tensions must be recognized and balanced rather than eliminated. Prior research has largely overlooked the role of paradoxes in ecosystem orchestration or has focused mainly on interfirm relationships. This study addresses this gap by examining a government led national QCI ecosystem that includes firms, research organizations, funding bodies, and governmental actors. Using an explorative case study with 15 informants from the Finnish QCI ecosystem and drawing on paradox theory as an analytical lens, we identify core paradoxical tensions and show how they challenge ecosystem orchestration. We contribute nuanced insights into the origins and dynamics of paradoxical tensions and discuss the implications for orchestrating multi-actor ecosystems.*

## 1    Introduction

The potential of quantum computing has sparked a global race, prompting nations to invest heavily in technology (Duranton, 2024). Although governments worldwide are developing their own national quantum ecosystems, the full-scale deployment of the technology still faces substantial challenges (Bova et al., 2023). To date, no actor has demonstrated quantum advantage, a practical computational advantage over classical systems (see Kwon et al., 2026), and it remains unclear which of the competing technologies will ultimately prevail.

This uncertainty also manifests at the ecosystem level, where nations are still experimenting with how to organize their quantum computing innovation (QCI) ecosystems. QCI refers to the development of quantum computing technologies, including advances in hardware, software, and algorithms, as well as their applications across domains. At present, national QCI ecosystems are in a formative phase, in which actors seek to orchestrate and integrate activities across diverse domains, aiming to align and mobilize the resources and capabilities of multiple stakeholders toward shared goals (cf. Mann et al., 2022). While, for example, AI ecosystems are strongly industry-led, QCI ecosystems are still science-focused and governments have a strong role in advancing their development (OECD, 2025; Ahmed et al., 2023). These QCI ecosystems typically involve a wide variety of stakeholders, including academia, industry, and government (Kar et al., 2024), and technology development efforts are fragmented across multiple levels (e.g., hardware and software) (Kwon et al., 2026). The diversity of interests and goals among ecosystem actors often generates tensions in orchestration processes. Some of these tensions display paradoxical features comprising persistent contradictions that are difficult to manage within the ecosystem (Smith & Lewis, 2011).

From a pragmatic perspective, identifying and understanding these paradoxical situations is essential for ecosystem development. In the context of ecosystem research, apart from a few notable studies (e.g.,

Schmeiss et al., 2019; Wareham et al., 2014), the role of paradoxes in ecosystem orchestration remains largely understudied. The extant literature lacks an understanding of how paradoxical tensions influence orchestration in innovation ecosystems. Moreover, prior research has largely overlooked stakeholders outside the B2B domain, and there is little, if any, literature on QCI orchestration. Consequently, the paradoxes influencing ecosystem orchestration warrant greater scholarly attention. To address this gap in theory and QCI practice, we formed the following research question: *What paradoxical tensions emerge in the orchestration of a national QCI ecosystem?*

We address this question through an explorative case study of the Finnish QCI ecosystem using an interpretive qualitative research approach (Walsham, 1995), thereby contributing to the IS research discourse. The Finnish QCI ecosystem aims to become industry-driven, but it is currently strongly government-led. The state plays a central role by funding research institutions and start-ups, acting as an early customer for quantum technologies, and investing in some quantum companies. Specifically, we synthesize the literature on ecosystem orchestration and paradoxical tensions to offer an integrated perspective that advances theoretical understanding and informs the practical management of emerging QCI ecosystems. Our preliminary findings offer valuable insights into what paradoxes QCI ecosystems contain and how these affect ecosystem orchestration.

## 2 Theoretical Background

### 2.1 Ecosystem Orchestration

In response to discontinuous technological change, organizations are increasingly compelled to form alliances to co-create value (Pushpananthan & Elmquist, 2022). Ecosystems enable this by facilitating information exchange and access to critical resources (Lusch & Nambisan, 2015). Despite the growing prominence of ecosystems, fewer than 15% have proven sustainable in the long term (Reeves et al., 2019). A major source of this fragility lies in the orchestration of the ecosystem (Pidun et al., 2021), which involves identifying, coordinating, and integrating the interdependent roles, capabilities, and actions of actors to achieve desired outcomes (Mann et al., 2022).

Effective orchestration enables ecosystems to create and distribute value, manage risks, and sustain competitive advantage (Pidun et al., 2021). Yet, ecosystem orchestration is complex and differs from other forms of interorganizational governance, as actors contribute unique and complementary capabilities (Lingens et al., 2022). Instead of traditional customer–supplier relationships, ecosystems rely on the voluntary collaboration of independent actors (Pidun et al., 2021). Because these actors depend on one another's core competencies to co-create value (Linde et al., 2021), the orchestrator cannot exert hierarchical control but must persuade partners to participate and cooperate. Successful orchestration therefore requires an understanding of the factors that shape collaboration, innovation, and value proposition design (Lingens et al., 2022). As actors pursue their own agendas and returns (Jacobides et al., 2018), paradoxes may arise, complicating orchestration (Mukhopadhyay & Bouwman, 2019).

Ecosystem orchestration has been examined especially in high-growth industry sectors, such as information technology (e.g., Zeng et al., 2023), as well as among firms of different sizes and stages of development (e.g., Lingens et al., 2021). Studies often focus on specific ecosystems led by industry leaders, such as Apple, Google, and Tencent. Likewise, much research has explored the diverse roles and responsibilities of ecosystem orchestrators (e.g., Mann et al., 2022). However, these studies mainly focus on ecosystems composed primarily of firms, while cross-sector ecosystems, especially those containing governmental actors, have received far less attention. Similarly, research on how orchestration unfolds in ecosystems involving multiple orchestrators is scarce (Lingens et al., 2022), particularly when the orchestrator operates outside the corporate domain. Moreover, there remains a limited understanding of the factors that give rise to tensions that complicate orchestration. This study addresses these gaps by examining the paradoxical tensions that influence orchestration within a cross-sector ecosystem. The focus is on the national QCI ecosystem, in which orchestration does not occur under the leadership of a single focal firm.

## 2.2 Paradox Theory and Ecosystems

The simultaneous need for stability and evolvability in technology ecosystems creates a paradox for their orchestration (Wareham et al., 2014). Paradox can be defined as "contradictory yet interrelated elements that exist simultaneously and persist over time" (Smith & Lewis, 2011, p. 382). This definition underscores the two fundamental elements of paradoxes: (1) the underlying tensions that give rise to them and (2) the actors' simultaneous engagement with these tensions (Smith & Lewis, 2011). Paradoxes entail managing inherent tensions through dynamic balance, turning contradictions into potential drivers of performance (see, e.g., Elo et al., 2024). This is why paradox theory provides a strong meta-theoretical lens for exploring the constructs, relationships, and dynamics of surrounding organizational tensions (Schad et al., 2016).

Paradox scholars have focused primarily on identifying types of paradoxes and their outcomes, paying less attention to the relationships within paradoxes and their dynamics (Schad et al., 2016). Previous studies have also focused mainly on intraorganizational tensions leading to paradoxes (Elo et al., 2024; Smania et al., 2024). This study responds to the gap highlighted by Tóth et al. (2022), emphasizing the need to broaden paradox theory categories to encompass interorganizational aspects of change. We address this research gap by examining orchestration challenges within a multisector ecosystem through the lens of paradox theory.

The paradox perspective suggests that sustaining long-term success depends on continuously addressing diverse and contradictory demands (Smith & Lewis, 2011). As organizations operate in ever more global and competitive contexts and as conflicting demands escalate, the paradox lens has emerged as a useful framework for understanding them (Smith & Lewis, 2011). Especially in the ecosystem context, identifying paradoxical tensions forms a critical foundation for orchestration (Wareham et al., 2014). Smith and Lewis (2011) identified paradoxes related to organizing, performing, belonging and learning. Organizing tensions stem from contradictions in structures and processes as organizations strive to balance competing operational demands. Performing tensions occur when actors pursue and promote competing goals. Belonging tensions arise from identity conflicts, including clashes of values and roles within actors. Finally, learning tensions emerge as organizations evolve, requiring them to draw on existing knowledge and capabilities while simultaneously exploring new possibilities. We adopt these four categories of paradoxical tensions as a lens to guide our exploration of the paradoxes behind multisector ecosystem orchestration.

## 3 Explorative Case Study Method

This study adopts an explorative case study design using an interpretive qualitative approach, focusing on a single case: the Finnish QCI ecosystem. A well-conducted and carefully documented interpretive case study can provide significant contributions to both IS theory and practice (Walsham, 1995). Such an approach is particularly suitable for investigating complex phenomena in real-life contexts because it deepens understanding of emerging issues by addressing the "why," "how," and "when" questions (Barrett & Walsham, 2004). The single case study approach was chosen for this research for two reasons. First, it enables us to explore the QCI ecosystem and its orchestration as it unfolds naturally. Second, it offers valuable insights into the paradoxical tensions arising across organizational boundaries within the QCI ecosystem.

The selected case, the Finnish QCI ecosystem, was chosen because of its timely relevance and access to rich, multidimensional data. At the study outset, Finland was aiming to become a major power in quantum technology and identified the quantum ecosystem as one of its greatest strengths. A publicly available draft of Finland's quantum strategy provided a valuable foundation for designing the semi-structured interviews. The Finnish QCI is especially interesting due to strong government involvement in funding quantum computer development rooted in decades of basic research[1]. The Finnish unicorn startup IQM Computers[2] has successfully developed and sold 15 commercial quantum computers, which range from 5 qubit research-purpose computers to 150 qubit versions for high-performance computing solutions. Algorithmiq[3] is pioneering the future of healthcare and life sciences by creating quantum algorithms that aim to harness the potential of quantum computing, and Bluefors[4] is the world-leading manufacturer of cooling solutions that are vital for quantum technology.

---

[1] See the Finnish Quantum Flagship research consortium for details, https://instituteq.fi

We interviewed 15 diverse QCI ecosystem actors, including representatives from government, research institutions, funding agencies, and private companies. Our goal was to investigate the QCI ecosystem dynamics, and power relations between actors. Semi-structured interviews included questions such as "How would you describe the ecosystem?", "How would you characterize the interactions between actors?" and "How is the ecosystem orchestrated?" The interviews were conducted remotely, audio-recorded, and later transcribed. The interviewees held senior leadership as well as expert positions within their organizations, for example in ministries and private firms. We summarize the details of the informants in Table 1. The informant selection followed a purposive sampling strategy (Patton, 2002), focusing on those organizations and individuals likeliest to offer valuable insights into the phenomenon under study. Thus, the participants differed in organizational structures, expectations, resources, and capabilities—and competed, particularly for resources. This created a paradox-rich context, offering an interesting setting for deeper examination. The data also included observations from four different events connected to the QCI ecosystem.

Thematic analysis (Terry et al., 2017) was used, and the data coding was guided by our theoretical pre-understanding and conceptual framework with four paradoxes (organizing, performing, belonging and learning). The data were examined inductively with the aim of identifying opinions that articulated divergent interpretations of the same phenomenon within the ecosystem. For example, regarding collaboration, first-order codes were assigned to statements indicating that collaboration functions well in the ecosystem, after which the material was systematically revisited to identify contrasting views on the same theme. When comments emphasizing mutual competition (i.e., oppositional views) appeared alongside positive accounts of collaboration, a tension was recorded. These tensions and their frequencies were then compiled and categorized under the four paradoxes. Coding was discontinued once no additional themes were identified in which the interviewees articulated conflicting views, indicating thematic saturation. The first author carried out the coding and analysis. To enhance the reliability of the findings (Nowell et al., 2017), the coding decisions and interpretations were subsequently discussed and critically reviewed in joint meetings with two of the authors.

| Sector | Industry | Informant role |
|---|---|---|
| Governmental actor (5) | Education, ICT, Cyber, Labor, Defense | Director of the ministry, Cyber security director, Research manager, Chief specialist, Senior specialist |
| Company (4) | Hardware (2), Software & hardware, Consulting | Government relations manager, Managing director, Quantum ambassador, Chief development officer |
| Research institution (3) | Technical Research Center (2), University | Research manager, Professor, Managing director |
| Funder (2) | State-owned (2) | Director of field, Lead specialist |
| Lobbyist (1) | Technology | Director |

*Table 1.    Sectors, industries, and roles of the interviewed QCI ecosystem actors.*

---

[2] https://meetiqm.com
[3] https://algorithmiq.fi/
[4] https://bluefors.com/

# 4    Preliminary Findings

In this section, we apply the four paradox categories introduced by Smith and Lewis (2011) to examine the tensions and resulting paradoxes emerging within the QCI ecosystem (Figure 1). The findings serve as a roadmap for the ecosystem orchestrator, supporting three key tasks: identifying, coordinating, and integrating (Mann et al., 2022). Our findings show how the ecosystem's paradoxical tensions jointly affect all three tasks of the orchestrator. First, the orchestrator must be aware of these tensions before they can be influenced. Second, tensions do not occur in isolation but interact with one another, which accentuates the importance of the integration task. Third, successful coordination of the ecosystem presupposes an understanding of the overall field of paradoxical tensions. Based on our results, however, the QCI ecosystem orchestrator can navigate by attending to which tensions occur more often than others. In our study, we focus on the "identifying" phase that aligns with the state of the Finnish QCI ecosystem while also offering preliminary insights into how the orchestrator can balance these tensions.

The Finnish QCI ecosystem has exhibited numerous tensions that have led to paradoxical situations requiring the orchestrator's attention. Most tensions emerged within the organizing category. The interviewees emphasized the importance of collaboration: "The collaboration is rather fragmented. Through cooperation, we could achieve more" (Interview 7, Governmental actor). At the same time, competition for resources such as talent and investments became evident, as one interviewee told us: "The shortage of skilled professionals is what everyone is facing" (Interview 13, Research institution), while another noted: "Research institutions and companies are also competing within the ecosystem" (Interview 2, Funder). Another significant tension concerns the governance of the ecosystem. It is perceived as currently weak, and there is no shared vision regarding its desired form. Some interviewees wished to preserve the ecosystem's self-organizing nature, while others called for stronger guidance: "In my opinion, it should probably be coordinated by the government" (Interview 3, Company); "Some kind of coordinator is definitely needed" (Interview 13, Research institution).

When looking at performance, there is clear tension between hardware and software within the ecosystem. Success in hardware development is viewed as a prerequisite for realizing the potential of quantum applications. At the same time, the software domain is perceived as offering greater long-term opportunities, suggesting that it should receive increased attention and investment: "I see a risk in our ecosystem that we may be focusing too much on the technology that is currently dominant" (Interview 15, Company). The QCI ecosystem is perceived to require a shared vision and collective efforts to advance it. However, this aspiration is challenged by the presence of numerous start-ups that lack the capacity to contribute meaningfully. One interviewee emphasized the situation as follows: "What currently slows down the ecosystem is that companies often lack the resources to engage in building a shared vision or common positions, as they have to focus primarily on survival" (Interview 3, Company). This is also reflected in the fact that the hardware side receives most attention: "The software side has somewhat been in the position of a child within the ecosystem" (Interview 2, Funder).

Under the learning category, the distinctive nature and strategic significance of quantum technology are emphasized for both companies and the state. The Finnish QCI ecosystem is seeking to expand its overall competence base, as the technology is viewed as critical to national interests: "This relates to the ecosystem, that we would have domestic encryption technology that is quantum-resistant" (Interview 7, Governmental actor). Because quantum technology encompasses multiple domains, such as quantum computing and photonics (Duranton, 2024), achieving expertise becomes challenging. The available knowledge cannot cover all areas, yet focusing solely on application development, which may yield higher profits in the future, is also seen as problematic. As one interviewee summarized: "A question is: Could we just sit and wait for IBM to build the quantum computer and then buy it? The argument is that for the development of the ecosystem (and knowledge), it is crucial at this stage that we also participate in the hardware development" (Interview 8, Research institution).

Finally, tensions closely linked to orchestration emerged under the belonging category. Power relations within the ecosystem are especially interesting due to the multifaceted role of the state. The government plays a central role in funding quantum technology, as it also steers the activities of funding agencies, a fact recognized also within the ecosystem: "Power lies where the money is" (Interview 15, Company).

However, an ecosystem sustained solely by public investments is not viewed as healthy: "In the end, it cannot endlessly rely on public investments, on the state subsidizing their research by providing the necessary conditions" (Interview 6, Funder). At the same time, the state is seen as an important first customer, a so-called "godfather"—a figure expected to share risks and provide early demand. Through the state-owned research institute VTT[5], the government has indeed acted as a key purchaser by acquiring a quantum computer from a company within the ecosystem. Thus, the state simultaneously seeks to remain neutral and to promote the ecosystem's development, yet also competes with the ecosystem actors. This multifaceted role poses a particular challenge for ecosystem orchestration.

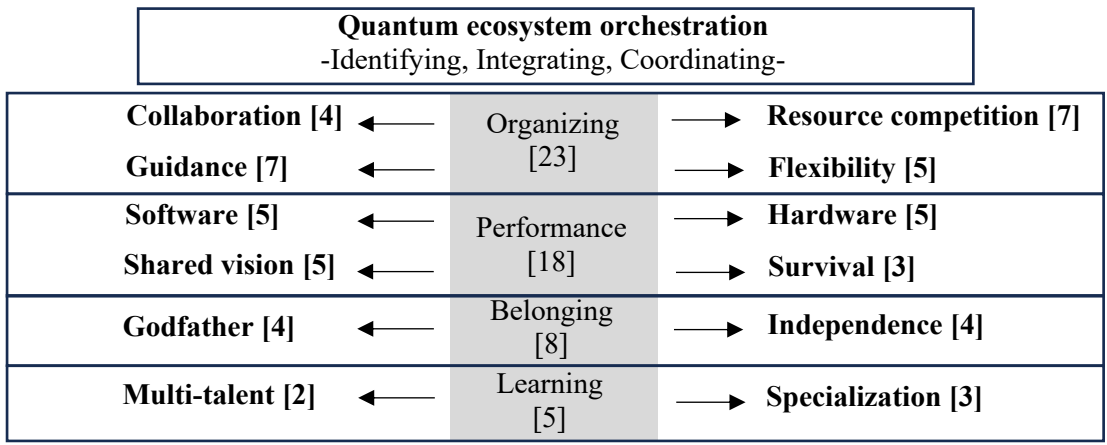

*Figure 1.    Paradoxes in Finnish QCI ecosystem orchestration. Note: The number in brackets represents how frequently observations of that specific tension occurred in the data.*

## 5    Discussion and Future Research Agenda

A successful ecosystem orchestrator requires a fundamental mindset shift from an inward-looking orientation to an outward-looking orientation (Mann et al., 2022). Achieving this shift presupposes identifying tensions that emerge within the ecosystem. This study examined the paradoxical tensions shaping the orchestration of the Finnish QCI ecosystem. Prior research has paid limited attention to the role of paradoxes in ecosystem orchestration or focused mainly on inter-firm relations. We addressed this gap by investigating paradoxical tensions in a multisector QCI ecosystem, where government plays a unique "godfather" role. Beyond identifying these tensions, our analysis explains how and why they manifest, interact and challenge orchestration.

Using paradox theory (Smith & Lewis, 2011) as an analytical lens, we identified six paradoxical tensions that affect the three key activities of ecosystem orchestrators—identifying, integrating, and coordinating (Mann et al., 2022). The Finnish QCI ecosystem is still emerging and seeking more industry-driven orchestration. Our results suggest that an orchestrator is likely to confront, in particular, collaboration–resource competition and guidance–flexibility-related paradoxical tensions. Software–hardware and shared vision–survival tensions also appear salient. Recognizing these paradoxes is critical for fostering a shared vision, balancing coopetition, and allocating scarce resources more effectively. Tensions in the learning and belonging categories are likewise important, as neglecting them may hinder capability-building and knowledge development within the ecosystem. Moreover, given the central role of the government in the ecosystem's evolution, orchestrator must be attentive to how this power asymmetry shapes the overall configuration and dynamics.

---

[5] https://www.vttresearch.com

Our findings resonate with earlier work on orchestration and paradoxical tensions (Smania et al., 2024; Tóth et al., 2022), revealing shared tensions related, for example, to goal divergence and coopetition. At the same time, we contribute nuanced insights into the mechanisms through which these tensions emerge and also identify new QCI-specific tensions. The uncertainty related to the future of quantum technology gives rise to tensions and "betting on the wrong horse" (e.g., software vs. hardware) may have expensive and far-reaching consequences. This uncertainty interlinks to human capital, creating a paradox around how best to deploy the ecosystem's scarce talent (multi-talent–specialization) in a field where the talent is scarce and grows slowly. More than half of quantum start-up founders hold a PhD, compared with around 10% of founders of companies in ecosystems in general (OECD, 2025). This intensifies the multi-talent–specialization tension. An important observation is that these paradoxes are closely interconnected and should not be addressed in isolation. For example, balancing the multi-talent-specialization paradox is difficult without simultaneously determining which technologies investments should be directed toward (hardware vs. software paradox). In addition, we show the paradox that in a nascent ecosystem, achieving a shared vision is difficult when many firms are still struggling for survival and searching for capabilities and profits. This paradox is closely interlinked with the guidance-flexibility paradox, as low levels of guidance create uncertainty among actors about whether their own actions and those of others really contribute to the common good. If others' actions do not appear to advance the collective good, actors may shift to prioritize their own survival.

Our study also reveals novel findings regarding the Finnish government's unique and paradoxical role in the QCI ecosystem. While the government cannot directly intervene in market competition, due the European Union regulation, and foster an industry-driven ecosystem, it can remain actively involved as a funder, supporter of scientific research, and early customer of quantum technologies. This "godfather"-role is unusual compared to other information technology ecosystems and interlinks other tensions. If the state were to significantly reduce its investments in quantum technology, competition over resources among ecosystem actors would likely intensify (collaboration vs. resource competition). Moreover, without a clearly communicated long-term governmental commitment, achieving a shared vision becomes difficult and smaller actors may increasingly focus on survival (shared vision vs. survival). Finally, the deep expertise required in quantum technology develops slowly and cannot realistically be sustained in academia without strong public support. Without strong support from the government, the ecosystem is forced to head towards limited resources (multi-talent vs. specialization).

For quantum computing practice and policy, our study provides ecosystem orchestrators (especially the government) with actionable insights into the opposing forces that arise in QCI ecosystems. Orchestrators can use this understanding to foster collaboration, direct resources more strategically, and ultimately enhance ecosystem performance. The government can use its influence to bring ecosystem actors together and create opportunities for collaboration, thereby balancing collaboration and resource competition. The national quantum strategies and roadmaps are tools which can be used to balance survival-behavior and foster shared vision. As an early customer, government can also steer companies' investments toward areas it considers strategically important, for example, national security (hardware vs. software). Because the QCI ecosystem is still strongly science-focused, the government can further influence which kind of future talent is trained by defining funding rules that fit to its purposes (multi-talent vs. specialization). However, if the QCI ecosystem is to become truly industry-driven, the government should gradually transfer power to other actors to avoid long-term dependency on the state (godfather vs. independence). This makes the "godfather" paradox even more important to balance.

It is important to acknowledge that paradoxical tensions are deeply interrelated and thus jointly influence all three orchestration activities. These tensions must first be identified before they can be constructively engaged; only then is it possible for orchestrators to integrate and coordinate actors and resources. We recognize that, from the orchestrator's perspective, the Finnish QCI ecosystem is still in the "identifying" phase. In future research, we aim to link other orchestrator functions (integrating and coordinating) to paradox theory. Through this approach, we can more comprehensively understand the effects of tensions within the QCI ecosystem and develop a deeper conceptual framework connecting paradox theory with ecosystem orchestration. Our research continues with further interviews across all segments of the QCI ecosystem and deepening of our analysis using the dataset of 22 public statements on Finland's quantum strategy and its associated QCI ecosystem, simultaneously triangulating these insights with the observations gathered. Our goal is to deepen the understanding of how these tensions dynamically interact and to focus especially on the government's unique steering role. We will further investigate how this "godfather"- role influences the QCI ecosystem power dynamics and how other actors' expectations toward it evolve. So far, we have focused on the paradoxical tensions in the Finnish QCI ecosystem. Going forward, we see it as essential to conduct comparative analyses with QCI ecosystems in other countries. Such comparisons could provide valuable insights into the paradoxes and orchestration practices of ecosystems at different stages of development and with various organizational structures. This would allow us to refine and validate our theoretical understanding of how paradoxical tensions evolve over time in a rapidly developing QCI ecosystem and how they shape the dynamics of ecosystem orchestration.

## Contact information

- Jori Taipale, University of Jyväskylä, Jyväskylä, Finland, jori.p.taipale@jyu.fi
- Olli Tyrväinen, University of Jyväskylä, Jyväskylä, Finland, olli.p.tyrvainen@jyu.fi
- Tuure Tuunanen, University of Jyväskylä, Jyväskylä, Finland, tuure.t.tuunanen@jyu.fi
- Joel Mero, University of Jyväskylä, Jyväskylä, Finland, joel.j.mero@jyu.fi
- Teiko Heinosaari, University of Jyväskylä, Jyväskylä, Finland, teiko.heinosaari@jyu.fi